\documentclass{mn2e}

\usepackage{graphicx,amsmath,amssymb,subfigure}

\def\simgt{\mathrel{\lower0.6ex\hbox{$\buildrel {\textstyle >}
 \over {\scriptstyle \sim}$}}}
\def\simlt{\mathrel{\lower0.6ex\hbox{$\buildrel {\textstyle <}
 \over {\scriptstyle \sim}$}}}
\newcommand{\etal}{{et al.}}                   

\newcommand{\Msolar}{\mbox{\,$\rm M_{\odot}$}}        

\begin{document}

\title[The discovery of a typical radio galaxy at $z = 4.88$]{The discovery of a typical radio galaxy at $z = 4.88$}

\author[Jarvis et al.]
{Matt J.~Jarvis$^{1}$\thanks{Email: M.J.Jarvis@herts.ac.uk}, Hanifa Teimourian$^{1}$, Chris Simpson$^{2}$, Daniel J.B.~Smith$^{2,3}$ \and Steve Rawlings$^{4}$ and David Bonfield$^{1}$\\
\footnotesize
$^{1}$Centre for Astrophysics, Science \& Technology Research Institute, University of Hertfordshire, Hatfield, Herts, AL10 9AB, UK\\
$^{2}$Astrophysics Research Institute, Liverpool John Moores University, Twelve Quays House, Egerton Wharf, Birkenhead, CH41 1LD, UK\\ 
$^{3}$School of Physics and Astronomy, University of Nottingham, University Park, Nottingham, NG72RD, UK\\
$^{4}$Astrophysics, Department of Physics, Keble Road, Oxford, OX1 3RH, UK \\ 
}

\maketitle

\begin{abstract}

In this letter we report the discovery of a $z=4.88$ radio galaxy discovered with a new technique which does not rely on pre-selection of a sample based on radio properties such as steep-spectral index or small angular size. This radio galaxy was discovered in the Elais-N2 field and has a spectral index of $\alpha = 0.75$, i.e. not ultra-steep spectrum. It also has a luminosity consistent with being drawn from the break of the radio luminosity function and can therefore be considered as a typical radio galaxy.
Using the {\em Spitzer}-SWIRE data over this field we find that the host galaxy is consistent with being similarly massive to the lower redshift powerful radio galaxies ($\sim 1-3L^{\star}$). We note however, that at $z=4.88$ the H$\alpha$ line is redshifted into the IRAC 3.6$\mu$m filter and some of the flux in this band may be due to this rather than stellar continuum emission. The discovery of such a distant radio source from our initial spectroscopic observations demonstrate the promise of our survey for finding the most distant radio sources. 

\end{abstract}


\section{Introduction}\label{sec:intro}

One of the major objectives in extragalactic astronomy today is to pin down
the Epoch of Reionization (EoR), where the Dark Ages comes to an end and UV photons are able to propagate throughout the Universe and ionize neutral hydrogen. Although the process of reionization
is expected to occur over a relatively short time interval (Gnedin
2000), the H{\sc i} absorption seen in the spectra of high-redshift
quasars (Fan et al.\ 2006) suggests that $z\sim6$ marks the threshold
of the EoR, while the 5-year WMAP data (Dunkley et al.\ 2009) suggest that it may have begun at a much earlier time, $z=11.0\pm1.4$. 

The first
opportunity to make detailed studies of the neutral hydrogen in the EoR is almost upon us with the imminent commissioning
of the Low Frequency Array (LOFAR) and the Murchison Widefield Array (MWA). LOFAR and the MWA will be able to detect the neutral hydrogen 21~cm transition line at redshifts beyond $z\sim 6$, predominantly searching for signatures via statistical fluctuations in the power spectrum (e.g. Iliev et al. 2008). However, another avenue for direct measurement of the neutral hydrogen fraction in the early Universe comes from observing the 21~cm forest against a bright distant source. This is also one of the key science questions for the Square Kilometre Array (SKA; see e.g. Carilli et al. 2002).

However, such experiments
obviously require a population of $z>6$ radio-loud sources against which
the 21\,cm absorption can be viewed and there is still much
uncertainty over the number of such sources. 
There have been many surveys to find very
high-redshift radio galaxies (HzRGs; e.g.  De Breuck et al. 2000, 2002; Jarvis et al. 2001a,b; Best et al. 2003; De Breuck et al. 2004; Jarvis et al. 2004; Cruz et al. 2006, 2007; Brookes et al. 2008) yet the highest redshift radio
galaxy known ``only'' lies at $z=5.19$ (van Breugel et al. 1999). See Miley \& De Breuck (2008) for a review on the various properties of distant radio galaxies.

This has led to a number of authors proposing that there is a ``cut-off'' in the radio source population at high redshift (e.g. Dunlop \& Peacock 1990; Shaver et al. 1996; Wall et al. 2005). However, analyses by other authors (e.g. Jarvis \& Rawlings 2000; Jarvis et al. 2001c) suggest that the sharpest reported declines may be the result of a range of selection effects and/or assumptions about the brightness of radio galaxy hosts (i.e. the $K-z$ relation; Lilly \& Longair 1984) holding to redshifts beyond those where there are any constraints.

Tracing the space-density evolution of these powerful sources would have significant implications for the build-up of the relation between the mass of a galaxy bulge and its central supermassive black hole (Magorrian et al. 1998) given the AGN-feedback prescriptions in semi-analytic models of galaxy formation (Bower et al. 2006; Croton et al. 2006). As the quasar nucleus is obscured in radio galaxies they provide us with a clear view of the stellar emission from the host galaxy.
With past searches for the HzRGs it has become apparent that powerful radio sources trace the most massive galaxies over the history of the Universe (e.g. Jarvis et al. 2001b). 

They have therefore been used as beacons to massive galaxies and (proto-)clusters at early times allowing the build-up of structure to be traced (e.g. Venemans et al. 2005; Miley et al. 2006). Furthermore, the HzRGs themselves may also be crucial in determining the impact that powerful radio activity has on the larger scale environments (Rawlings \& Jarvis 2004; Gopal-Krishna \& Wiita 2001). Thus, our survey not only aims to produce the first sample of distant radio sources for investigating the EoR but also provide a much larger sample for investigations of the formation and evolution of the most massive galaxies at the highest redshifts.

In this letter we report the discovery of the second most distant radio galaxy, found in preliminary observations from our new, highly efficient survey to search for the most distant radio sources.

Throughout this letter the AB magnitude
system is used (Oke \& Gunn, 1983), and a standard cosmology is
assumed in which $H_{0}$ = 72 km s$^{-1}$, $\Omega_{M}$ = 0.26 and
$\Omega_{\Lambda}$ = 0.74 (Dunkley et al. 2009).

\section{Our Survey}\label{sec:survey}

HzRGs (by which we mean those at $z>4$) represent only a very small fraction of all radio sources at a given
flux-density limit and it is necessary to filter out low-redshift contaminants
before performing spectroscopy. This has previously been done by applying a filter based on radio
properties such as steep spectral index (e.g. Chambers et al. 1996; Blundell et al. 1998; De Breuck et al, 2002; Cohen et al. 2004; Cruz et al. 2006). The next stage
has typically been to take $K$-band images of the targets, since radio galaxies follow
an extremely tight locus in the $K$-band Hubble diagram (Jarvis et al. 2001b; Willott et al; 2003; Bryant et al. 2009) and therefore HzRGs will be faint. 
However,
although $K$-band imaging is very reliable at identifying radio
galaxies, these observations are very expensive, needing to reach
$K\sim20$ to detect sources at the epoch where the AGN density peaks
($z\sim2$; Jarvis \& Rawlings 2000; Willott et al. 2001). Unfortunately, even after filtering on radio source properties, such as angular size and spectral index, the fraction of
HzRGs remains low -- e.g., 1/68 sources in the 6C** sample of Cruz et
al.\ (2007) lies at $z>4$ -- so most of the follow-up near-infrared imaging 
simply rules out high-redshift targets. We are now in a position to turn this technique around and use large radio surveys over patches of sky which have relatively deep and wide near-infrared data.

The recent advent of wide-field deep
near-infrared surveys has opened up a new path for finding such objects. The
{\em Spitzer}-SWIRE (Lonsdale et al. 2003) and the UKIDSS Deep Extragalactic Survey (DXS; see e.g. Warren et al. 2007) are deep enough to reliably
eliminate virtually all $z<2$ radio galaxies. Therefore by
targeting only radio sources which are very faint or not detected in the {\em Spitzer}-IRAC channels 1 and 2 or the
DXS $K$-band imaging, we are likely to be probing radio sources in the 3\,Gyr after the Big Bang.

We have cross-matched the {\em Spitzer}-SWIRE data with sources at $>10$mJy from the 1.4~GHz FIRST survey (Becker et al. 1995) in the Lockman Hole, Elais-N1 and Elais-N2 fields, which together cover a total of $\sim 24$~square degrees.  For our initial search we use a 3.6$\mu$m limit of $<30\mu$Jy in a 1.9~arcsec radius aperture, as given in the SWIRE catalogues. 
Our choice of radio flux-density limit ensures that all objects at $z>2$
are above the break in the radio luminosity function and will
therefore possess strong emission lines, particularly Ly$\alpha$ which is redshifted into the optical part of the spectrum, allowing redshift identification in much shorter integration times than if signal-to-noise on the continuum is needed.

\begin{figure}
\includegraphics[width=0.85\columnwidth]{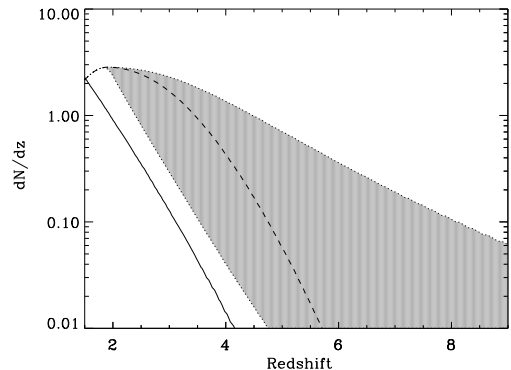}
\caption{
Expected number of radio sources above 10~mJy as a function of redshift per square degree. The dark solid line is the expected number of low-luminosity radio sources which would have low-luminosity emission lines and be very difficult to obtain redshifts for. The grey banded area shows the degree of uncertainty in the high-redshift evolution of the more powerful, typically FRII sources, from a constant comoving space density and one with a steady decline [see Jarvis \& Rawlings (2000) for more details]. The dashed line is the best-fit model C from Willott et al. (2001).}
\label{fig:dndz}
\end{figure}

Simulations (Jarvis \& Rawlings 2004; Wilman et al. 2008)
predict that $\sim$1 in 15 (adopting the average model; Fig.~\ref{fig:dndz}) of these sources will lie at $z>4$.
This flux-density limit and faint near-infrared magnitude cut also ensures that the number of lower-redshift lower luminosity radio sources are also reduced significantly (see e.g. Fig.~\ref{fig:dndz}) with only $\sim 1$ in 10  expected at $z>2$ in 24 square degrees. The survey will be described in more detail in a forthcoming paper (Teimourian et al. in prep.).

\section{A radio galaxy at $z=4.88$}\label{sec:results}

\begin{figure}
\includegraphics[width=0.85\columnwidth]{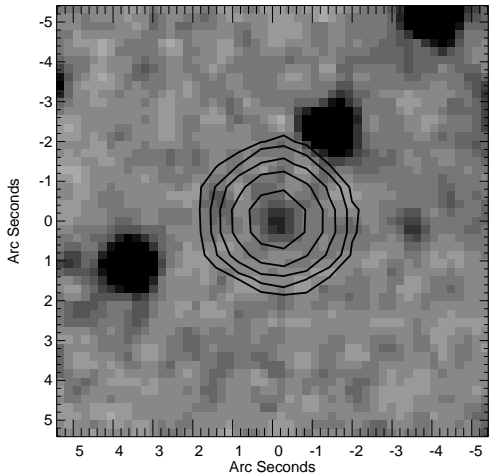}
\caption{The {\em Spitzer}-3.6$\mu$m image (greyscale) overlaid with radio contours from the FIRST survey. Contour levels are 0.8, 1.6, 3.2, 6.4 and 12.8~mJy/beam. One can see the faint source at the centre of the radio emission at 16 39 12.11 +40 52 36.5 which we identify as the host galaxy.}
\label{fig:overlay}
\end{figure}

J163912.11+405236.5 is detected in the FIRST survey with a flux-density of 22.5~mJy and is unresolved at the 5~arcsec resolution of this survey. This source is also identified in the Northern VLA Sky Survey (NVSS; Condon et al. 1998) with $S_{1.4 \rm GHz} = 21.8 \pm 0.8$~mJy and is therefore consistent with being point like. There is also a source detected in the 325~MHz Westerbork Northern Sky Survey (WENSS; Rengelink et al. 1997) at 16 39 12.17 +40 52 40.3 (J2000) which is 3.6~arcsec away from the FIRST centroid, therefore we associate this source with the FIRST source. The WENSS catalogue gives a flux-density of $67\pm5$~mJy for this source. Assuming a power-law spectral index between 325~MHz and 1.4GHz the spectral index of the source is therefore $\alpha = 0.75\pm0.05$\footnote{We use the convention for spectral index $S_{\nu} \propto \nu^{-\alpha}$}. Thus this source would not fall into the category of ultra-steep spectrum sources which have been used to search for high-redshift radio galaxies in recent years. In Fig.~\ref{fig:overlay} we show the 3.6$\mu$m image from the {\em Spitzer}-SWIRE survey overlaid with the radio image from the FIRST survey. There is a very faint source in the 3.6$\mu$m image at the centre of the radio position which we identify as the host galaxy. As this object is on the SWIRE survey area there are also a wealth of imaging data from the INT at optical wavelengths which allow us to constrain the continuum emission from this object. We also use the SWIRE data to constrain the brightness of the source from 3.6-24$\mu$m. Photometry of the host galaxy is presented in Table~\ref{tab:photometry}.

\subsection{The Lyman-$\alpha$ emission}

We observed the radio source J163912.11+405236.5 on the 27th April 2009 with the ISIS spectrograph on the William Herschel Telescope. We used the standard 5300\AA\, dichroic to reflect the blue light to an EEV12  $4096\times2048$ pixel$^{2}$ CCD using the R300B grism. The red light was dispersed through the R158R grism on to RED+ $4096\times2048$ pixel$^{2}$  CCD. We used a slit width of 2~arcsec which gave a resolution of 8.6\AA\, in the blue arm and 16.5\AA\, in the red arm.

The data were bias subtracted, illumination corrected and flat-fielded using standard IRAF tasks. Wavelength calibration was carried out using CuNe+CuAr lamps and flux-calibration was done using the spectrophotometric standard Feige 34.

\begin{figure}
\includegraphics[width=0.85\columnwidth]{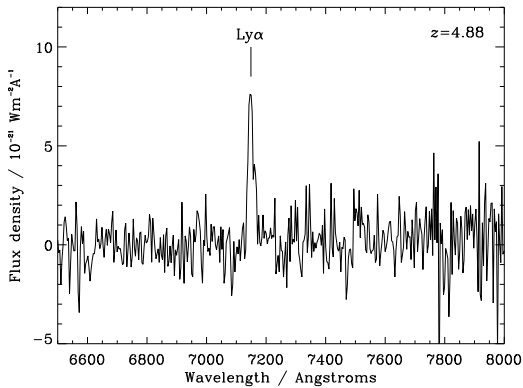}
\caption{
1-dimensional spectrum of the radio galaxy  J163912.11+405236.5. One can see the bright Lyman-$\alpha$ emission line at $\lambda=7149$\AA, which corresponds to a redshift of $z=4.88$. }
\label{fig:spectrum}
\end{figure}

In Fig.~\ref{fig:spectrum} we show the 1D-spectrum of the radio galaxy  J163912.11+405236.5 which displays a strong emission line at 7149\AA, which we identify as Lyman-$\alpha$ due to its blue cut-off and broader red wing, at a redshift of $z=4.88$. The line has a flux of $1.85\times10^{-19}$~W~m$^{-2}$, which at this redshift corresponds to a luminosity of $L_{{\rm Ly} \alpha} = 4.7\times10^{36}$~W and a rest-frame full-width half maximum of 1040~km~s$^{-1}$, i.e. a narrow-line AGN.
There are no other emission lines in our spectra which also suggest that the line is indeed Lyman-$\alpha$ rather than other common emission lines in radio galaxies. However, one possibility is that the emission line is  [OII]$\lambda3727$ at $z=0.92$. If this were indeed [OII] from a narrow-line AGN then we would expect to see the CII]$\lambda2326$ and the CIII]$\lambda1909$ emission line in the blue end of the spectrum, for which we find no evidence, in addition to a much brighter host galaxy. Another possibility is that it is a starburst galaxy. Using the relation of Kewley, Geller \& Jansen (2004) the [OII] luminosity would correspond to a star-formation rate of $SFR \sim 11~\Msolar$~yr$^{-1}$. The radio emission one would expect from this starburst galaxy is greater than three orders of magnitudes fainter than what we observe. We are therefore confident that the line is Ly$\alpha$ at $z = 4.88$.

This is the second highest redshift radio galaxy known to date and demonstrates the promise of our new survey in finding high-redshift radio sources. Given the redshift, the radio luminosity of the source at 325~MHz is $\log_{10}(L_{325\rm MHz} /~$W~Hz$^{-1}$~sr$^{-1}$) = 26.94 and $\log_{10}(L_{1400\rm MHz} /~$W~Hz$^{-1}$~sr$^{-1}$) = 26.47 at 1.4~GHz. This is very close to the break luminosity in the radio luminosity function of Willott et al (2001) which is where the bulk of the luminosity density arises, and can therefore be classed as a typical radio source at this redshift. We also see hints of associated absorption redward of the line centre. This is reminiscent of the associated absorption found in other powerful radio galaxies (van Ojik et al. 1997; Jarvis et al. 2003; Wilman et al. 2004) and, given the compact nature of the radio source, may also reinforce the idea that small and young HzRGs may be surrounded by a shell of neutral gas which is linked to the fueling of the AGN (e.g. Emonts et al. 2007), although higher resolution spectroscopy is needed to confirm this.

\begin{table}
\centering
\caption{\label{tab:photometry}Photometric data for the radio galaxy.The quoted limits are 3$\sigma$ limits for a 2~arcsec diameter aperture for the INT data, 4~arcsec diameter aperture for the channels 1-4 IRAC data and a 10.5~arcsec diameter aperture for the 24$\mu$m limit. All are in the AB magnitude system. The radio galaxy is unresolved in both the FIRST and WENSS survey data and as such the radio flux densities are integrated flux over the synthesized beam area. }
\begin{tabular*}{5.19cm}{|c|c|}
\hline
Photometric Band & AB Magnitude        \\
\hline
u'       & $>24.8$\\
g'         & $>25.5$   \\
r'     &    $>25.1$ \\
i'         &    $>24.1$\\
z' & $>23.2$\\
3.6$\mu$m       &   $21.97 \pm 0.13$  \\
4.5$\mu$m       &   $>21.7$  \\
5.8$\mu$m & $>20.0$\\
8.0$\mu$m & $>19.9$\\ 
24$\mu$m   &  $>18.1$          \\
\hline
325 MHz          &  $67 \pm 5$~mJy        \\
1.4 GHz          &  $22.5 \pm 0.4$~mJy         \\
\hline
\end{tabular*}
\end{table}

\subsection{The host galaxy}\label{host}

At $z = 4.88$ the IRAC channels on {\em Spitzer} sample longward of the 4000\AA\, break and the emission is presumably dominated by an old stellar population. 
Using the flux-density at 3.6$\mu$m and assuming a $K-3.6\mu$m colour of 2.25 from the models of Bruzual \& Charlot (2003), we find that the host galaxy would have a $K-$band magnitude of $K_{AB}=24.2$. Using $K_{AB} - K_{Vega} = 1.9$, we are able to determine where this source would lie on the radio galaxy $K-z$ relation. Using the relation from Willott et al. (2003) for $z=4.88$, we find that the expected $K-$band magnitude is $K_{Vega} = 20.34$, whereas we find a $K-$band magnitude of $K_{Vega} = 22.3$. Thus this is around 2 magnitudes fainter than the $K-z$ relation determined at low redshift. This is not unexpected as the 4000\AA\, break lies between the $K$ and 3.6$\mu$m bands at $z = 4.88$, thus if there is little ongoing star formation in the host galaxy it would be extremely difficult to detect at $K-$band even if the host galaxy is relatively massive. If we compare the measured 3.6$\mu$m flux with the radio galaxies in the study of Seymour et al. (2007) the only source which is at comparable redshift, and where the 4000\AA\, break would be redshifted beyond the $K-$band, is TN~J0924-2201 at $z=5.195$. This has a reported flux density of 11.3 $\pm$ 1.8$\mu$Jy at 3.6$\mu$m, a factor of two brighter than  J163912.11+405236.5. TN~J0924-2201 has a radio luminosity that is approximately 1.5~dex brighter than our source, thus this is possibly related to the fact that the brightest radio galaxies seem to reside in more massive host galaxies (e.g. Eales et al. 1997; Willott et al. 2003; McLure et al. 2004), both of which may be related to the central supermassive black hole. Unfortunately the lack of a detection at the longer wavelength IRAC channels precludes us from carrying out SED fits to estimate the mass of the object. Furthermore, at $z=4.88$ the H$\alpha$ line falls within the 3.6$\mu$m filter and we cannot be sure whether the emission is indeed due to the stellar continuum emission or a strong emission line. If we assume case B recombination we expect the flux ratio Ly$\alpha$/H$\alpha  = 8.7$ (Brocklehurst 1971) which would then lead to a magnitude in the 3.6$\mu$m band of $AB \sim 26$, therefore this does seem unlikely. However, much lower flux ratios have been found in radio galaxies (see e.g. McCarthy et al. 1993) and our measured Ly$\alpha$ flux is possibly an underestimate of the true emission if there is indeed a large amount of associated absorption and/or obscuration by dust. Therefore, we cannot completely rule out significant line emission in the 3.6$\mu$m band from H$\alpha$.


\section{Conclusions}\label{sec:conclusions}

We have devised a new survey to find the most distant radio sources with the aim of detecting a source suitable for 21~cm absorption studies with the LOFAR. Initial spectroscopy of this survey has resulted in the discovery of the second most distant radio galaxy known at a redshift of $z = 4.88$. This source has a spectral index, measured between 325~MHz and 1.4~GHz, of $\alpha=0.75$ and therefore does not belong to the class of ultra-steep spectrum sources, samples of which have been used to find the majority of the highest redshift radio sources thus far (but see Waddington et al. 1999). 

Its luminosity is very close to the break luminosity in the radio luminosity function of Willott et al. (2001) and can therefore be described as a typical radio galaxy at these early epochs. This differs from the other known high-redshift radio galaxies, which are generally towards the extreme luminosities and are thus considerably rarer than the more typical population.

Using the IRAC photometry from the SWIRE survey we are able to determine where this radio galaxy would lie on the $K-z$ relation, and find that it is approximately 2~magnitudes fainter than that inferred from the $K-z$ relation if one uses a simple extrapolation of the relation found by Willott et al. (2003). This can be explained by the fact that the 4000\AA\,break is redshifted beyond the $K-$band and as such is no longer sampling the bulk of the stellar emission from older stars. We therefore conclude that it is possible that this high-redshift radio galaxy is similarly massive to radio galaxies at lower redshifts. We note however that the 3.6$\mu$m filter would contain the H$\alpha$ emission line at this redshift and the detected emission could be at least partially a result of this.

The discovery of such a distant radio source from our initial spectroscopic observations demonstrate the promise of our survey for finding the most distant radio sources.
We did not observe all of our candidates HzRGs in this observing run due to poor weather, as a consequence  the effective area surveyed was only $\sim$4.5~square degrees. 
With further observations over the rest of the SWIRE and UKIDSS DXS regions, we will certainly be able to make the most accurate measurements of the space density of radio sources at $z>3$ and hopefully discover the first radio source appropriate for 21~cm absorption studies within the epoch of reionization. This discovery also shows that samples constructed on the basis of steep-spectral index may miss a significant fraction of high-redshift sources (see also Waddington et al. 1999).
Future observations of the rest of the $S_{1.4\rm GHz}>10$~mJy sources in the SWIRE/DXS fields will allow us to quantify the distribution of spectral indices for these distant sources and aid in defining the search criteria for distant radio sources with future radio continuum surveys made possible with LOFAR, EVLA and the SKA precursor telescopes.
Furthermore, future observations with the VISTA Deep Extragalactic Observations (VIDEO) survey (http://star-www.herts.ac.uk/$\sim$mjarvis/video/) and the Spitzer Representative Volume survey (SERVs; http://www.its.caltech.edu/$\sim$mlacy/servs.html) will not only allow to make an extremely selective sample of HzRG candidates but also allow us to measure the environmental density around these distant radio sources in the same way that has been attempted around the most distant QSOs (e.g. Willott et al. 2005; Stiavelli et al. 2005).

\section*{ACKNOWLEDGEMENTS} 
MJJ acknowledges the support of an RCUK fellowship.
 The WHT is operated on the island
of La Palma by the Isaac Newton Group in the Spanish Observatorio del
Roque de los Muchachos of the Instituto de Astrofisica de
Canarias.

{} 

\end{document}